\documentstyle[epsfig,12pt,a4p]{article}

\newcommand{\kk}{\mbox{$K^+K^-$ }}
\newcommand{\ksks}{\mbox{$K^0_SK^0_S$ }}

\newcommand{\pipi}{\mbox{$\pi^+\pi^-$ }}
\newcommand{\pipizer}{\mbox{$\pi^0\pi^0$ }}

\begin{document}
\begin{titlepage}
\def\footnoterule{\hrule width 1.0\columnwidth}
\begin{tabbing}
put this on the right hand corner using tabbing so it looks
 and neat and in \= \kill
\> {27 July 1999}
\end{tabbing}
\bigskip
\bigskip
\begin{center}{\Large  {\bf A coupled channel analysis of the centrally
produced
$K^+ K^-$ and $\pi^+\pi^-$ final states
in $pp$ interactions at 450~GeV/c}
}\end{center}
\bigskip
\bigskip
\begin{center}{        The WA102 Collaboration
}\end{center}\bigskip
\begin{center}{
D.\thinspace Barberis$^{  4}$,
F.G.\thinspace Binon$^{   6}$,
F.E.\thinspace Close$^{  3,4}$,
K.M.\thinspace Danielsen$^{ 11}$,
S.V.\thinspace Donskov$^{  5}$,
B.C.\thinspace Earl$^{  3}$,
D.\thinspace Evans$^{  3}$,
B.R.\thinspace French$^{  4}$,
T.\thinspace Hino$^{ 12}$,
S.\thinspace Inaba$^{   8}$,
A.\thinspace Jacholkowski$^{   4}$,
T.\thinspace Jacobsen$^{  11}$,
G.V.\thinspace Khaustov$^{  5}$,
J.B.\thinspace Kinson$^{   3}$,
A.\thinspace Kirk$^{   3}$,
A.A.\thinspace Kondashov$^{  5}$,
A.A.\thinspace Lednev$^{  5}$,
V.\thinspace Lenti$^{  4}$,
I.\thinspace Minashvili$^{   7}$,
J.P.\thinspace Peigneux$^{  1}$,
V.\thinspace Romanovsky$^{   7}$,
N.\thinspace Russakovich$^{   7}$,
A.\thinspace Semenov$^{   7}$,
P.M.\thinspace Shagin$^{  5}$,
H.\thinspace Shimizu$^{ 10}$,
A.V.\thinspace Singovsky$^{ 1,5}$,
A.\thinspace Sobol$^{   5}$,
M.\thinspace Stassinaki$^{   2}$,
J.P.\thinspace Stroot$^{  6}$,
K.\thinspace Takamatsu$^{ 9}$,
T.\thinspace Tsuru$^{   8}$,
O.\thinspace Villalobos Baillie$^{   3}$,
M.F.\thinspace Votruba$^{   3}$.
Y.\thinspace Yasu$^{   8}$.
}\end{center}

\begin{center}{\bf {{\bf Abstract}}}\end{center}

{
A coupled channel analysis of the centrally produced \kk and \pipi
final states has been performed in $pp$ collisions at
an incident beam momentum of 450~GeV/c.
The pole positions and branching ratios to $\pi\pi$ and $K \overline K$
of the
$f_0(980)$, $f_0(1370)$, $f_0(1500)$ and $f_0(1710)$ have been determined.
A systematic study of the production properties of all the resonances
observed in the \pipi and \kk channels has been performed.
}
\bigskip
\bigskip
\bigskip
\bigskip\begin{center}{{Submitted to Physics Letters}}
\end{center}
\bigskip
\bigskip
\begin{tabbing}
aba \=   \kill
$^1$ \> \small
LAPP-IN2P3, Annecy, France. \\
$^2$ \> \small
Athens University, Physics Department, Athens, Greece. \\
$^3$ \> \small
School of Physics and Astronomy, University of Birmingham, Birmingham, U.K. \\
$^4$ \> \small
CERN - European Organization for Nuclear Research, Geneva, Switzerland. \\
$^5$ \> \small
IHEP, Protvino, Russia. \\
$^6$ \> \small
IISN, Belgium. \\
$^7$ \> \small
JINR, Dubna, Russia. \\
$^8$ \> \small
High Energy Accelerator Research Organization (KEK), Tsukuba, Ibaraki 305-0801,
Japan. \\
$^{9}$ \> \small
Faculty of Engineering, Miyazaki University, Miyazaki 889-2192, Japan. \\
$^{10}$ \> \small
RCNP, Osaka University, Ibaraki, Osaka 567-0047, Japan. \\
$^{11}$ \> \small
Oslo University, Oslo, Norway. \\
$^{12}$ \> \small
Faculty of Science, Tohoku University, Aoba-ku, Sendai 980-8577, Japan. \\
\end{tabbing}
\end{titlepage}
\setcounter{page}{2}
\bigskip
\par
Recently the WA102 collaboration has published the results of
partial wave analyses of the centrally produced \kk, \ksks~\cite{kkpap},
\pipi~\cite{pipipap} and \pipizer~\cite{pi0pi0pap} channels.
A striking feature of these analyses was the result that the $f_J(1710)$
has J~=~0 (we shall refer to it as the $f_0(1710)$ hereafter).
In these papers the S-wave from each channel was fitted independently
using interfering Breit-Wigners and a background.
In this present paper we will first show how the resulting parameters change
if a different method of fitting is used, namely, a T-Matrix analysis and
a K-Matrix analysis using the methods described in ref.~\cite{bugg}.
Next we will perform a coupled channel fit to the
\kk and \pipi final states in order to determine the
pole positions and branching ratios of the observed mesons. Finally we will
present information on the production kinematics of these
resonances.
\par
In our previous publication a fit has been performed to the \pipi S-wave using
a coherent sum of relativistic Breit-Wigner functions and a background of the
form:
\[
A(M_{\pi\pi}) = Bgd(M_{\pi\pi}) + \sum_{n=1}^{N_{res}}
a_ne^{i\theta_n}BW_n(M_{\pi\pi})
\label{eq:a}
\]
where the background has been parameterised as
\[
Bgd(M_{\pi\pi}) = \alpha (M_{\pi\pi} - 2m_\pi)^\beta e^{-\gamma M_{\pi\pi} -
\delta
M_{\pi\pi}^2}
\label{eq:b}
\]
where $a_n$ and $\theta_n$ are the amplitude and the phase of the $n$-th
resonance respectively,
$\alpha$, $\beta$, $\gamma$ and $\delta$
are real parameters, $BW(M_{\pi\pi})$ is the
relativistic
Breit-Wigner function for a spin zero resonance. In order to describe the
centrally produced
\pipi mass spectrum the function $|A(M_{\pi\pi})|^2$ has been multiplied by the
kinematical factor $(M_{\pi\pi} - 4m_\pi^2)^{1/2}/M_{\pi\pi}^3$~\cite{AMP}.
The resulting function is then convoluted with a Gaussian to account for the
experimental mass
resolution.
\par
In this present paper we use the Flatt\'{e}
formula~\cite{flatte}
to describe the $f_0(980)$, this is referred to as Method I.
For the \pipi channel the Breit-Wigner has the
form:
\[
BW(M_{\pi\pi}) = \frac{m_0 \sqrt{\Gamma_i} \sqrt{\Gamma_\pi}}
{m_0^2 - m^2 -im_0(\Gamma_\pi + \Gamma_K)}
\]
and in the \kk channel the Breit-Wigner has the form:
\[
BW(M_{KK}) = \frac{m_0 \sqrt{\Gamma_i} \sqrt{\Gamma_K}}
{m_0^2 - m^2 -im_0(\Gamma_\pi + \Gamma_K)}
\]
where $\Gamma_i$ is absorbed into the intensity of the resonance.
$\Gamma_\pi$ and
$\Gamma_K$ describe the partial widths of the resonance to
decay to $\pi \pi$ and $K \overline K$
and are given by
\[
\begin{array}{c}
\Gamma_\pi = g_\pi(m^2/4 - m^2_\pi)^{1/2} \\
\\
\Gamma_K =g_k/2 [(m^2/4 - m^2_{K^+})^{1/2} + (m^2/4-m^2_{K^0})^{1/2} ]
\end{array}
\]
where $g_\pi$ and $g_K$ are the squares of the coupling constants of the
resonance
to the $\pi \pi $ and $K \overline K$ systems.
The resulting fit is shown in fig.~\ref{fi:1}a) for the entire mass spectrum
and in fig.~\ref{fi:1}b) for masses above 1~GeV.
The sheet II pole positions~\cite{sheet} for the resonances are
\begin{tabbing}
00000aaaa\=adfsfsf99ba \=Mas  == 1224 pm0 \=i12\=2400 \=pi \=1200000  \=MeV
\kill
\>$f_0(980)$ \>M $ \; = \;$($\; \; 983$$\; \pm\; \; \;$8) \>$-i$\>($\;
\;58$\>$\pm$\>11)\>MeV \\
\>$f_0(1370)$ \>M $ \; = \;$(1306$\; \pm\; $18)\>$-i$\>(111\>$\pm$\>23)\>MeV\\
\>$f_0(1500)$ \>M $ \; = \;$(1502$\; \pm\; $12)\>$-i$\>($ \;
\;65$\>$\pm$\>12)\>MeV\\
\>$f_0(1710)$ \>M $ \; = \;$(1748$\; \pm\; $22)\>$-i$\>($ \;
\;73$\>$\pm$\>22)\>MeV
\end{tabbing}
These parameters are consistent with the PDG~\cite{PDG98} values for these
resonances.
\par
To test the sensitivity of these results on the fitting method used
we have also performed a fit to the \pipi mass spectrum using the
T-Matrix parameterisation of Zou and Bugg~\cite{bugg}.
The invariant amplitude for \pipi central production can be expressed as
\begin{equation}
A = \alpha_1(s)T_{11} + \alpha_2(s)T_{21}
\label{eq:c}
\end{equation}
where $T_{11}$ and $T_{21}$ are the invariant amplitudes for elastic $\pi\pi
\rightarrow \pi\pi$
and $K \overline K \rightarrow \pi\pi$ scattering and are parameterised by
\begin{equation}
T_{11} = \frac{e^{2i\phi} - 1}{2i\rho_1} + \frac{g_1e^{2i\phi}}{M_R^2 -s -
i(\rho_1g_1 + \rho_2g_2)}
\label{eq:d1}
\end{equation}
\begin{equation}
T_{21} = \frac{\sqrt{g_1g_2}e^{i\phi}}{M_R^2 - s - i(\rho_1g_1+\rho_2g_2)}
\label{eq:d2}
\end{equation}
where $\rho_1 = (1-4m_\pi^2/s)^{1/2}$ and
$\rho_2 = (1-4m_K^2/s)^{1/2}$ are phase space factors and $s$ is the invariant
mass squared of the
\pipi channel. The background term is presumed to be coupled only to the
$\pi\pi$ channel
and has the form
\[
T_b = \frac{e^{2i\phi} - 1}{2i\rho_1} = \frac{1}{A(s) - i\rho_1}
\]
which satisfies the unitarity condition. $A(s)$ is an arbitrary real function
which has been
taken to be of the form
\[
A(s)=\frac{1+a_1s+a_2s^2}{b_1(s-m_\pi^2/2)+b_2s^2}
\]
\par
The real functions $\alpha_i(s)$ in equation~(\ref{eq:c}) describe the coupling
of the
initial state to the channel $i$. These functions are approximated by the power
expression~\cite{AMP}:
\[
\alpha_i(s) = \sum_{n=0}\alpha_i^n\{\frac{s}{4m_K^2}\}^n
\]
where the factor $4m_K^2$ is introduced as a convenient scaling.
It has been found that $n$~=~3 is sufficient to
describe the S-wave distribution.
\par
In order to describe the centrally produced
\pipi mass spectrum,
the function $|A(M_{\pi\pi})|^2$ has been multiplied by the
kinematical factor $(M_{\pi\pi} - 4m_\pi^2)^{1/2}/M_{\pi\pi}^3$~\cite{AMP} and
the resulting function is then convoluted with a Gaussian to account for the
experimental mass
resolution.
The resulting fit is shown in fig.~\ref{fi:1}c) for the entire mass spectrum
and in fig.~\ref{fi:1}d) for masses above 1~GeV.
As can be seen the fit
does not
describe well the region above 1.0 GeV.
The sheet II pole corresponding to the $f_0(980)$ is
\begin{tabbing}
0000000aaaa\=MS \= = \=12240 \=pi \=120 \=i22\=2400 \=pi \=1200000  \=MeV
\kill
\>$M$ \>= \>($\; \;993$ \>$\pm$ \>$\; \;8$) \>$-i$\>($\; \;38$\>$\pm$\>$\;
\;9$) \>MeV
\end{tabbing}
There are two poles from the background term
with
\begin{tabbing}
0000000aaaa\=MS \= = \=12240 \=pi \=120 \=i22\=2400 \=pi \=1200000  \=MeV
\kill
\>$M_1$ \>= \>($\; \;388$ \>$\pm$ \>55) \>$-i$\>(223\>$\pm$\>28) \>MeV \\
\>$M_2$ \>= \>(1541 \>$\pm$ \>32) \>$-i$\>(143\>$\pm$\>21) \>MeV \\
\end{tabbing}
The first pole may be associated with the low mass S-wave.
The NA12/2 collaboration~\cite{NA122} have previously
shown that this region may be interpreted as being due to the $\sigma$
particle.
The second pole would
appear to be in the region of the $f_0(1500)$ but the width is very broad.
This is due to the fact that the fit is not able to properly
describe the
region around 1.3 GeV.
\par
Adding one more term to equations~(\ref{eq:d1}) and (\ref{eq:d2})
to describe the 1300~MeV mass region improves the fit considerably
but still does not describe the region around 1700~MeV.
In order
to produce a satisfactory fit
two terms have to be added to
equations~(\ref{eq:d1}) and (\ref{eq:d2})
to account for the $f_0(1370)$ and $f_0(1710)$
which results in the fit
shown
in fig.~\ref{fi:1}e) for masses above 1~GeV.
The new sheet II pole
positions
are
\begin{tabbing}
0000aaaa\=adfsfsf99ba \=Ma \= = \=122499 \=pm \=120 \=i22\=2400 \=pi \=1200000
\=MeV   \kill
\>$f_0(980)$ \>M \>=\>($\; \;992$\>$\pm$\>$\; \;6$)
\>$-i$\>($\;\;52$\>$\pm$\>$\;\;9$)\>MeV \\
\>$f_0(1370)$ \>M \>=\>(1310\>$\pm$\>30)\>$-i$\>(134\>$\pm$\>23)\>MeV\\
\>$f_0(1500)$ \>M \>=\>(1497\>$\pm$\>17)\>$-i$\>($\;\;82$\>$\pm$\>21)\>MeV\\
\>$f_0(1710)$ \>M \>=\>(1752\>$\pm$\>15)\>$-i$\>($\;\;53$\>$\pm$\>12)\>MeV
\end{tabbing}
These parameters are consistent with the values from the
fit using interfering Breit-Wigners and with the
PDG~\cite{PDG98} values for these resonances.
\par
An alternative parameterisation is to use the K-Matrix formalism.
In this case the
Lorentz invariant T-Matrix is expressed as
\[
\hat{T} = \hat{K}(1-i \hat{\rho} \hat{K})^{-1}
\]
where for the case of $\pi\pi$ and $K \overline K$ final states $\hat{\rho}$
is a 2 dimensional
diagonal matrix and $\hat{K}$ is a
real symmetric 2 dimensional matrix of the form
\[
K_{ij} = \frac{a_ia_j}{M_R^2 -s} + \frac{b_ib_j}{s_b-s} + \gamma_{ij}
\]
In the K-Matrix formalism the background is assumed to couple to both the
$\pi\pi$ and
$K \overline K$ channels.
In order to describe the centrally produced
\pipi mass spectrum,
the function $|A(M_{\pi\pi})|^2$ has been multiplied by the
kinematical factor $(M_{\pi\pi} - 4m_\pi^2)^{1/2}/M_{\pi\pi}^3$~\cite{AMP} and
the resulting function is then convoluted with a Gaussian to account for the
experimental mass resolution.
\par
Two coupled channel resonances are found from the fit
shown in fig.~\ref{fi:1}f) for the entire mass spectrum
and in fig.~\ref{fi:1}g) for masses above 1~GeV
with their sheet II T-Matrix poles at
\begin{tabbing}
0000aaaa\=MR \= = \=12244 \=pi \=120 \=i22\=2400 \=pi \=1200000  \=MeV   \kill
\>$M_1$ \>= \>($\; \;988$ \>$\pm$ \>18) \>$-i$\>($\;\;39$\>$\pm$\>12) \>MeV \\
\>$M_2$ \>= \>(1526 \>$\pm$ \>22) \>$-i$\>(191\>$\pm$\>53) \>MeV \\
\end{tabbing}
As in the case of the T-Matrix analysis the fit fails in the 1.3 GeV region.
Adding one additional pole improves the fit. However, in order to
to obtain a satisfactory fit it is found necessary to include
two extra poles.
The fit shown is in fig.~\ref{fi:1}h) for masses above 1~GeV and results in
sheet II pole positions of
\begin{tabbing}
0000aaaa\=adfsfsf99ba \=Ma \= = \=12249 \=pi \=120 \=i22\=2400 \=pi \=1200000
\=MeV   \kill
\>$f_0(980)$ \>M \>=\>($\;\;982$\>$\pm$\>$\;\;9$)
\>$-i$\>($\;\;38$\>$\pm$\>16)\>MeV \\
\>$f_0(1370)$ \>M \>=\>(1290\>$\pm$\>30)\>$-i$\>(104\>$\pm$\>25)\>MeV\\
\>$f_0(1500)$ \>M \>=\>(1510\>$\pm$\>10)\>$-i$\>($\;\;56$\>$\pm$\>15)\>MeV\\
\>$f_0(1710)$ \>M \>=\>(1709\>$\pm$\>15)\>$-i$\>($\;\;75$\>$\pm$\>18)\>MeV
\end{tabbing}
These parameters are consistent with the values from the two previous fits.
\par
Finally,
in order to perform a coupled channel fit to the \pipi and \kk final states
a correct normalisation of the two data sets has to be performed.
The fit has been modified to take into account
the relative differences in geometrical
acceptance, event reconstruction and event selection.
The fit also includes corrections for the unseen decay modes
so that the branching ratio $\pi\pi$ to $K \overline K$ can be calculated.
In addition to the resonances discussed above, the
$a_0(980)$ can also contribute to the $K \overline K$ S-wave mass spectrum.
The contribution from $a_0(980) \rightarrow K^+ K^-$ has been calculated
from the observed decay
$a_0(980) \rightarrow \eta \pi$ and the measured branching ratio
of the $a_0(980)$~\cite{f1pap}. The calculated contribution (500~$\pm$~120
events)
has been included in the fit as a histogram.
There is no evidence for any $a_0(1450)$ contribution
in the $\eta \pi$ channel
and hence it has not been included in the fit to the $K \overline K$ S-wave.
\par
The coupled channel fit has been performed using the three methods described
above. The following pole positions and branching ratios
quoted for the resonances are a mean from the
three methods. The statistical error
is the largest error from the three fits and the
systematic error quoted represents the spread of the
values from the three methods. The results of the combined fit are
shown in fig.~\ref{fi:2}.
The sheet II pole positions are
\begin{tabbing}
0000aaaa\=adfsfsf99ba \=Ma \= = \=12249 \=pi \=12000000 \=i22\=2400 \=pi
\=1200000000  \=MeV   \kill
\>$f_0(980)$ \>M \>=\>($\;\;987$\>$\pm$\>$\;\;6  \pm \; \;6$)
\>$-i$\>($\;\;48$\>$\pm$\>12  $\pm \;\;  8$)\>MeV \\
\>$f_0(1370)$ \>M \>=\>(1312\>$\pm$\>25  $\pm$ 10)\>$-i$\>(109\>$\pm$\>22
$\pm$  15)\>MeV\\
\>$f_0(1500)$ \>M \>=\>(1502\>$\pm$\>12  $\pm$
10)\>$-i$\>($\;\;49$\>$\pm$\>$\;\;9$  $\pm \;$  8)\>MeV\\
\>$f_0(1710)$ \>M \>=\>(1727\>$\pm$\>12  $\pm$
11)\>$-i$\>($\;\;63$\>$\pm$\>$\;\;8$  $\pm \;$  9)\>MeV
\end{tabbing}
These
parameters are consistent with the PDG~\cite{PDG98} values for these
resonances.
For the $f_0(980)$ the couplings were determined to be
$g_\pi$~=~0.19~$\pm$~0.03~$\pm$~0.04
and $g_K$~=0.40~$\pm$~0.04~$\pm$~0.04.
\par
The branching ratios for the $f_0(1370)$, $f_0(1500)$ and
$f_0(1710)$ have been calculated to be:
\[
\frac{f_0(1370) \rightarrow K \overline K}{f_0(1370) \rightarrow \pi \pi} =
0.46 \pm 0.15 \pm 0.11
\]
\[
\frac{f_0(1500) \rightarrow K \overline K}{f_0(1500) \rightarrow \pi \pi} =
0.33 \pm 0.03 \pm 0.07
\]
\[
\frac{f_0(1710) \rightarrow K \overline K}{f_0(1710) \rightarrow \pi \pi} = 5.0
 \pm 0.6 \pm 0.9
\]
These values are to be compared with the
PDG~\cite{PDG98} values of 1.35~$\pm$~0.68 for the $f_0(1370)$ and
0.19~$\pm$~0.07 for the $f_0(1500)$, which comes from the Crystal Barrel
experiment~\cite{cbf1500}.
The value for the $f_0(1710)$ is consistent with the value of
2.56~$\pm$~0.9  which comes from the WA76 experiment
which assumed J~=~2 for the $f_J(1710)$~\cite{oldkkpipi}.
\par
In our previous publications
a study has been performed of resonance production
rate
as a function of
the difference in the transverse momentum vectors ($dP_T$)
between the particles exchanged from the
fast and slow vertices~\cite{WADPT,closeak}.
It has been observed\cite{memoriam} that
all the undisputed
$ q \overline q $ states
(i.e. $\eta$, $\eta^{\prime}$, $f_1(1285)$ etc.)
are suppressed as $dP_T$ goes to zero,
whereas the glueball candidates
$f_0(1500)$ and $f_2(1950)$ survive.
In order to calculate the contribution of each resonance as a function
of $dP_T$, the partial waves have been fitted
in three $dP_T$ intervals with
the parameters of the resonances fixed to those obtained from the
fits to the total data using method I.
As an example of how the mass spectra change as a function of $dP_T$,
fig.~\ref{fi:3} shows the \pipi S-wave and D-wave in three
$dP_T$ intervals.
\par
The S-wave clearly shows that the region around 1.3 GeV is more enhanced
at large $dP_T$ in contrast to the low mass part of the spectrum.
This is not due to feed through from the D-wave which has been estimated to
be less
than 3~\% in this mass region.
In the D-wave the $f_2(1270)$ is suppressed at small $dP_T$ in contrast to the
behaviour of the low mass part of the D-wave.
\par
Table~\ref{ta:dpt} gives the percentage of each resonance
in three $dP_T$ intervals together with the ratio of the number of events
for $dP_T$ $<$ 0.2 GeV to
the number of events
for $dP_T$ $>$ 0.5 GeV for each resonance considered.
As can be seen from table~\ref{ta:dpt}, the $\rho^0(770)$, $f_2(1270)$ and
$f_2^\prime(1525)$ are suppressed at small $dP_T$ in contrast to the
$f_0(980)$, $f_0(1500)$ and $f_0(1710)$.
\par
The azimuthal angle ($\phi$) is defined as the angle between the $p_T$
vectors of the two protons.
In order to determine the
$\phi$ dependence for the resonances observed,
the partial waves have been fitted in 30 degree bins of $\phi$
with the parameters of the resonances fixed to those obtained from the
fits to the total data.
The fraction of each resonance as a function of $\phi$ is plotted in
fig.~\ref{fi:5}.
The $\phi$ dependences are clearly not flat and considerable variation
is observed between the different resonances.
\par
In order to determine the
four momentum transfer dependence ($t$) of the
resonances observed
in the \pipi and \kk channels
the partial waves have been fitted in 0.1 GeV$^2$ bins
of $t$
with the parameters of the resonances fixed to those obtained from the
fits to the total data.
Fig.~\ref{fi:6} shows the four momentum transfer from
one of the proton vertices for these resonances.
The distributions  for the
$f_0(980)$, $f_0(1370)$, $f_0(1500)$, $f_0(1710)$ and $\rho(770)$
have been fitted with a single exponential
of the form $exp(-b |t|)$ and the
values of $b$ found are given in table~\ref{ta:b}.
The distributions for the $f_2(1270)$ and $f_2^\prime(1525)$
cannot be fitted with a single
exponential.
Instead they have been fitted to the form
\[
\frac{d\sigma}{dt} = \alpha e^{-b_1t} + \beta t e^{-b_2t}
\]
The parameters resulting from the fit are
given in table~\ref{ta:c}.
\par
In a previous publication by the WA76 collaboration~\cite{oldpipi}
the ratios of the production
cross sections for the $\rho(770)$, $f_0(980)$ and $f_2(1270)$
were calculated
at $\sqrt s$~=~12.7 and 23.8~GeV.
However,
the experiment at 300 GeV ($\sqrt s$~=~23.8~GeV) was only sensitive to $\phi$
angles
less than 90 degrees and
the acceptance program that had been used assumed a flat $\phi$ distribution.
Hence the cross sections at 300 GeV were underestimated for the $\rho(770)$
and $f_2(1270)$ and overestimated for the $f_0(980)$.
\par
After correcting for
geometrical acceptances, detector efficiencies,
losses due to cuts,
and unseen decay modes,
the ratios of the cross-sections for the
$\rho(770)$, $f_0(980)$, $f_2(1270)$ and in addition, the $f_0(1500)$
at $\sqrt s$~=~29.1 and 12.7~GeV are given in table~\ref{ta:d}.
The cross sections for
the $f_0(980)$, $f_2(1270)$ and $f_0(1500)$ are compatible with
being independent of energy
which
is consistent with them being produced via double Pomeron exchange~\cite{dpet}.
\par
In summary, a coupled channel fit has been performed to the
centrally produced \pipi and \kk mass spectra. The pole positions
and branching ratios of the $f_0(980)$, $f_0(1370)$, $f_0(1500)$
and $f_0(1710)$ have been determined using three different methods
which give consistent results.
An analysis of the $dP_T$ dependence of the resonances observed
shows that the undisputed $q \overline q$ mesons are
suppressed at small $dP_T$ in contrast to the
enigmatic $f_0(980)$, $f_0(1500)$ and $f_0(1710)$.
Considerable variation is observed in the $\phi$ distributions
of the produced mesons.
\begin{center}
{\bf Acknowledgements}
\end{center}
\par
This work is supported, in part, by grants from
the British Particle Physics and Astronomy Research Council,
the British Royal Society,
the Ministry of Education, Science, Sports and Culture of Japan
(grants no. 04044159 and 07044098), the French Programme International
de Cooperation Scientifique (grant no. 576)
and
the Russian Foundation for Basic Research
(grants 96-15-96633 and 98-02-22032).

\newpage
\newpage
\begin{table}[h]
\caption{Production of the resonances as a function of $dP_T$
expressed as a percentage of their total contribution and the
ratio (R) of events produced at $dP_T$~$\leq$~0.2~GeV to the events
produced at $dP_T$~$\geq$~0.5~GeV.}
\label{ta:dpt}
\vspace{1in}
\begin{center}
\begin{tabular}{|c|c|c|c|c|} \hline
 & & & & \\
 &$dP_T$$\leq$0.2 GeV & 0.2$\leq$$dP_T$$\leq$0.5 GeV &$dP_T$$\geq$0.5 GeV &
$R=\frac{dP_T \leq 0.2 GeV}{dP_T\geq 0.5 GeV}$\\
 & & & & \\ \hline
 & & & & \\
$f_0(980)$  &23.1$\pm$ 2.2 & 50.6 $\pm$ 2.7   &26.2 $\pm$ 2.7 & 0.88~$\pm$~0.12
\\
 & & & & \\ \hline
 & & & & \\
$f_0(1370)$  &18.1 $\pm$ 4.0 & 32.0 $\pm$ 2.0  &49.9 $\pm$ 2.9  &
0.36~$\pm$~0.08\\
 & & & & \\ \hline
 & & & & \\
$f_0(1500)$  &23.5 $\pm$ 2.5  & 54.1 $\pm$ 5.0  &22.3 $\pm$ 3.0   &
1.05~$\pm$~0.18 \\
 & & & & \\ \hline
 & & & & \\
$f_0(1710)$  &26.4 $\pm$ 1.2  & 45.8 $\pm$ 1.0 &27.8 $\pm$ 1.1 &
0.95~$\pm$~0.06\\
 & & & & \\ \hline
 & & & & \\
$\rho(770)$  &6.4 $\pm$ 2.5  & 41.1 $\pm$ 4.0 &52.5 $\pm$ 3.0 &
0.12~$\pm$~0.05\\
 & & & & \\ \hline
 & & & & \\
$f_2(1270)$  &7.6 $\pm$ 1.2  & 29.5 $\pm$ 0.6 &62.9 $\pm$ 1.7 &
0.12~$\pm$~0.02\\
 & & & & \\ \hline
 & & & & \\
$f_2^\prime(1525)$  &4.3 $\pm$ 3.0  & 35.7 $\pm$ 3.0 &60.0 $\pm$ 4.0 &
0.07~$\pm$~0.04\\
 & & & & \\ \hline
 & & & & \\
$f_2(2150)$  &2.6 $\pm$ 3.0  & 53.6 $\pm$ 3.6 &43.6 $\pm$ 3.2 &
0.06~$\pm$~0.08\\
 & & & & \\ \hline
\end{tabular}
\end{center}
\end{table}
\newpage
\begin{table}[h]
\caption{The slope parameter $b$ from a single exponential fit to the $|t|$
distributions.}
\label{ta:b}
\vspace{0.3in}
\begin{center}
\begin{tabular}{|cccccc|} \hline
  &&&& & \\
 & $f_0(980)$ & $f_0(1370)$ & $f_0(1500)$ & $f_0(1710)$ & $\rho(770)$ \\
   &&&&& \\ \hline
  &&&& & \\
 b/GeV$^{-2}$ & $5.3 \pm 0.2$& $6.7 \pm 0.5$& $5.2 \pm 0.5$& $6.5 \pm 0.8$&
$5.8 \pm 0.1$ \\
   &&&&& \\ \hline
\end{tabular}
\end{center}
\end{table}
\begin{table}[h]
\caption{The slope parameters from a fit to the
$|t|$ distributions of the form
$\frac{d\sigma}{dt} = \alpha e^{-b_1t} + \beta t e^{-b_2t}$.}
\label{ta:c}
\vspace{0.3in}
\begin{center}
\begin{tabular}{|ccccc|} \hline
  &&&&  \\
 & $\alpha$ & $b_1$ & $\beta$ & $b_2$\\
&&GeV$^{-2}$&&GeV$^{-2}$ \\
   &&&& \\ \hline
  &&&& \\
$f_2(1270)$ & $0.25\pm 0.04$& $8.3 \pm 3.0$& $5.3 \pm 1.0$& $8.7 \pm 0.4$ \\
  &&&& \\
$f_2^\prime(1525)$ & $0.4\pm 0.3$& $8.3 \pm 5.0$& $5.3 \pm 4.0$& $7.7 \pm 3.3$
\\
   &&&& \\ \hline
\end{tabular}
\end{center}
\end{table}
\begin{table}[h]
\caption{The ratio of the cross sections at $\sqrt s$~=~29.1 and 12.7~GeV.}
\label{ta:d}
\vspace{0.3in}
\begin{center}
\begin{tabular}{|lcccc|} \hline
  &&& & \\
 & $\rho(770)$ & $f_0(980)$ & $f_2(1270)$ & $f_0(1500)$ \\
   &&&& \\ \hline
  &&& & \\
 $\frac{\sigma(\sqrt{s}=29.1)}{\sigma(\sqrt{s}=12.7)}$ & $0.36 \pm 0.05$& $1.28
\pm 0.21$& $0.98 \pm 0.13$& $1.07 \pm 0.14$ \\
   &&&& \\ \hline
\end{tabular}
\end{center}
\end{table}
\clearpage
{ \large \bf Figures \rm}
\begin{figure}[h]
\caption{The \pipi mass spectrum with a) and b) fit using
interfering Breit-Wigners.
Fit using the T-Matrix parameterisation c) and d) the original fit and
e) including
two extra poles.
Fit using the K-Matrix parameterisation f) and g) the original fit
and h) including
another two poles.
}
\label{fi:1}
\end{figure}
\begin{figure}[h]
\caption{ A coupled channel fit to the \pipi and \kk S-wave distributions.
}
\label{fi:2}
\end{figure}
\begin{figure}[h]
\caption{The \pipi S-wave and D-wave in three $dP_T$ intervals.}
\label{fi:3}
\end{figure}
\begin{figure}[h]
\caption{The $\phi$ distributions of the resonances observed in the
\pipi and \kk channels.}
\label{fi:5}
\end{figure}
\begin{figure}[h]
\caption{The four momentum transfer squared ($|t|$) from one of the proton
vertices
for the resonances observed in the
\pipi and \kk channels.}
\label{fi:6}
\end{figure}
\newpage
\begin{center}
\epsfig{figure=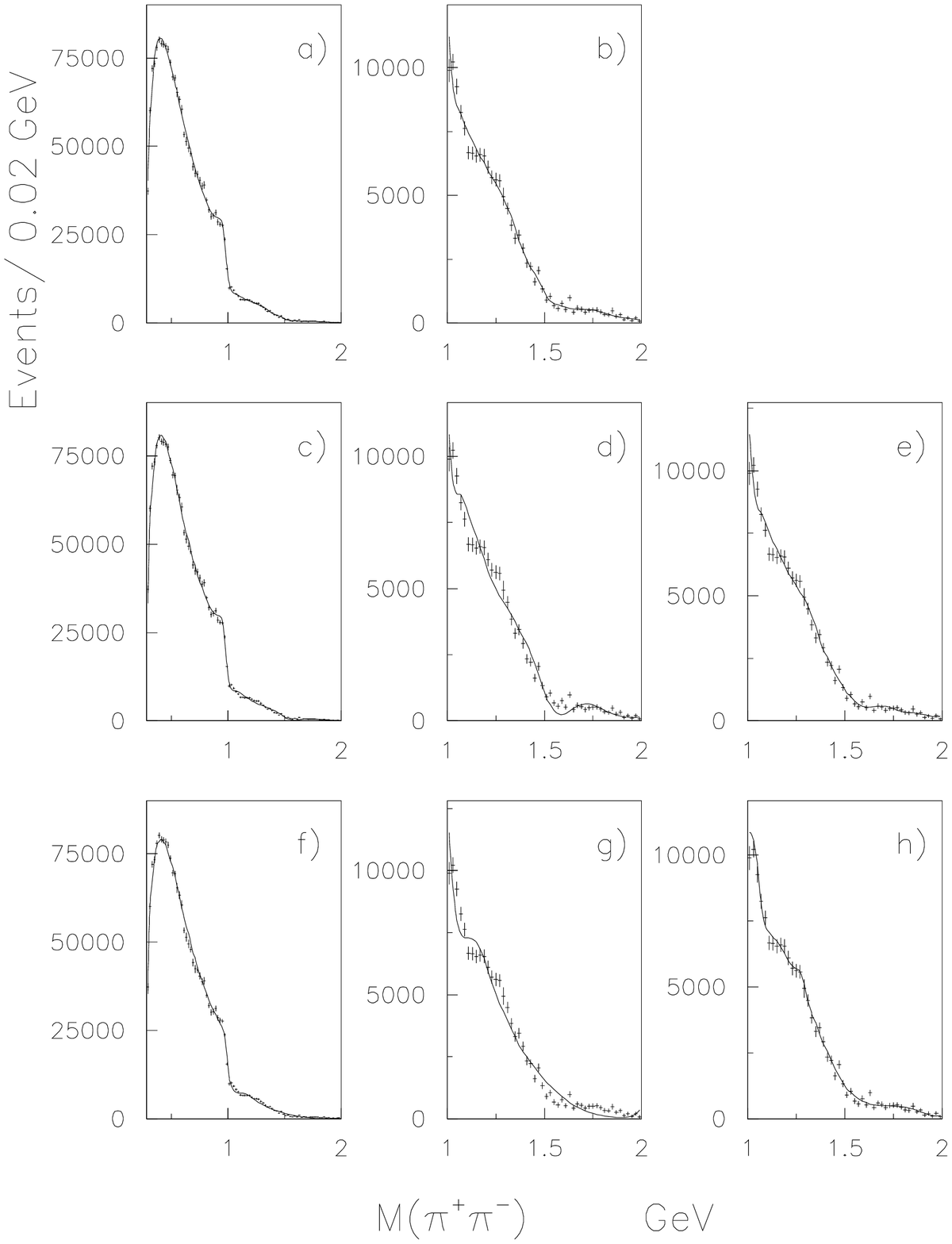,height=22cm,width=17cm}
\end{center}
\begin{center} {Figure 1} \end{center}
\newpage
\begin{center}
\epsfig{figure=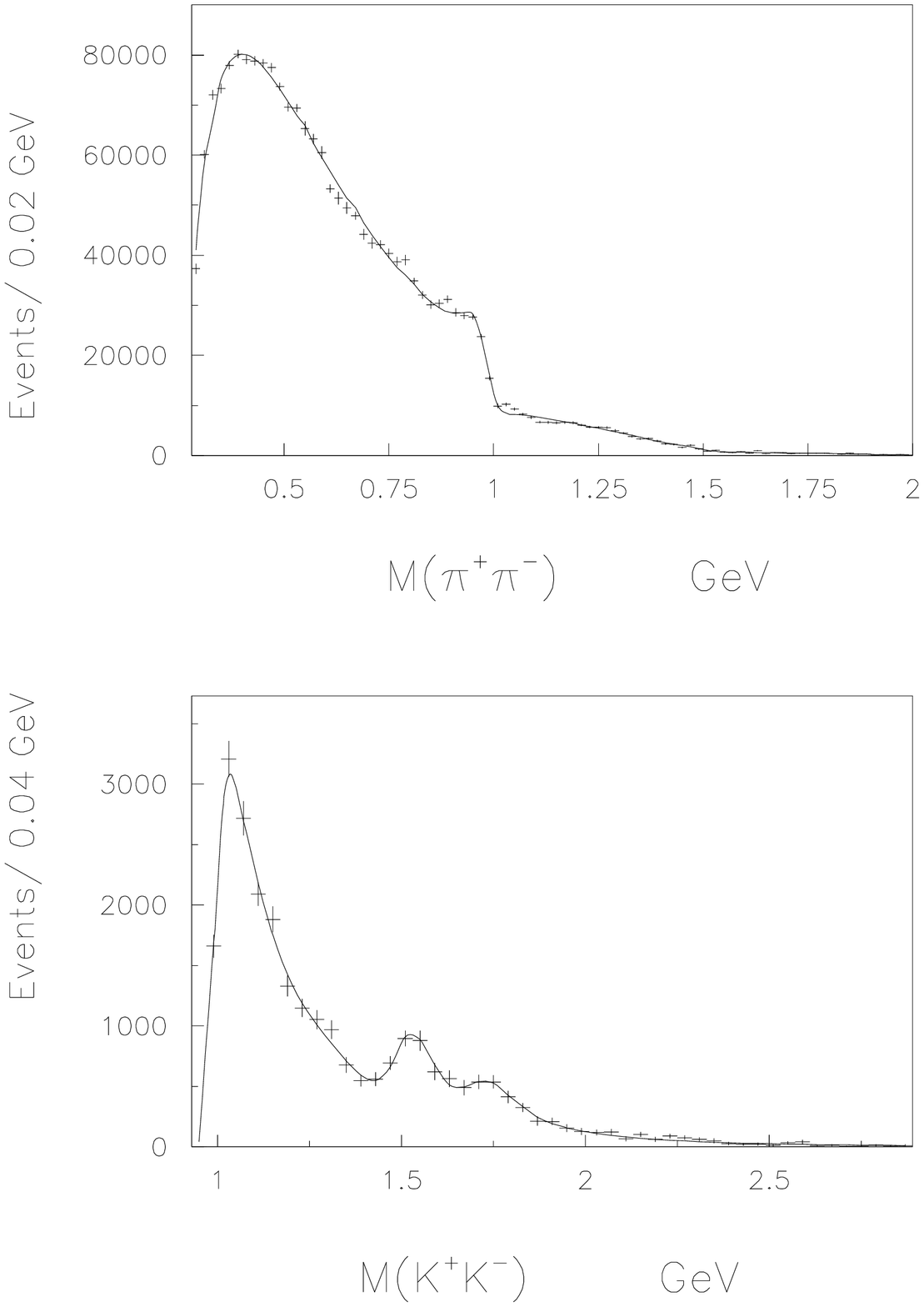,height=22cm,width=17cm,
bbllx=0pt,bblly=0pt,bburx=550pt,bbury=657pt}
\end{center}
\begin{center} {Figure 2} \end{center}
\newpage
\begin{center}
\epsfig{figure=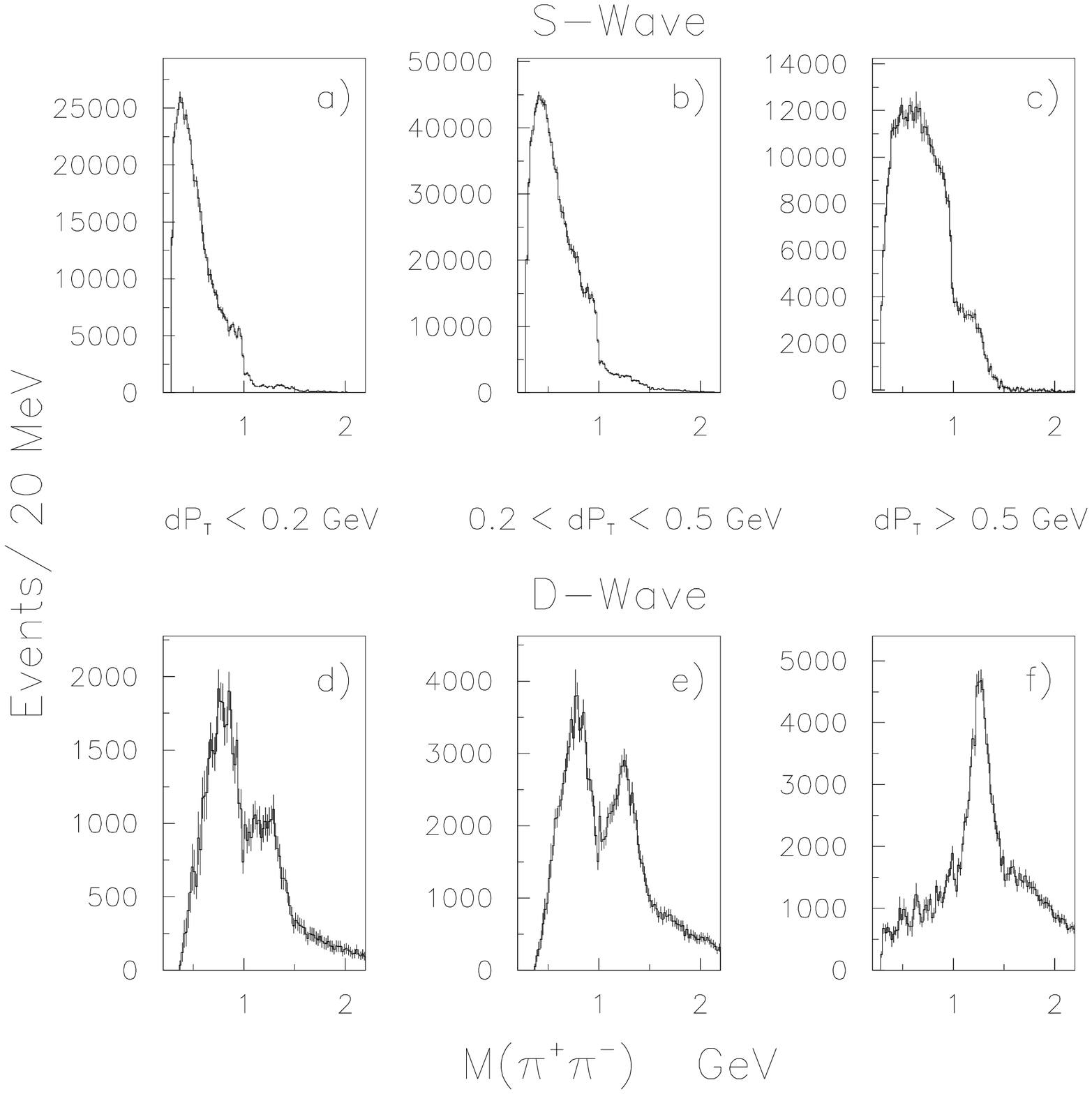,height=22cm,width=17cm,
bbllx=0pt,bblly=0pt,bburx=650pt,bbury=750pt}
\end{center}
\begin{center} {Figure 3} \end{center}
\newpage
\begin{center}
\epsfig{figure=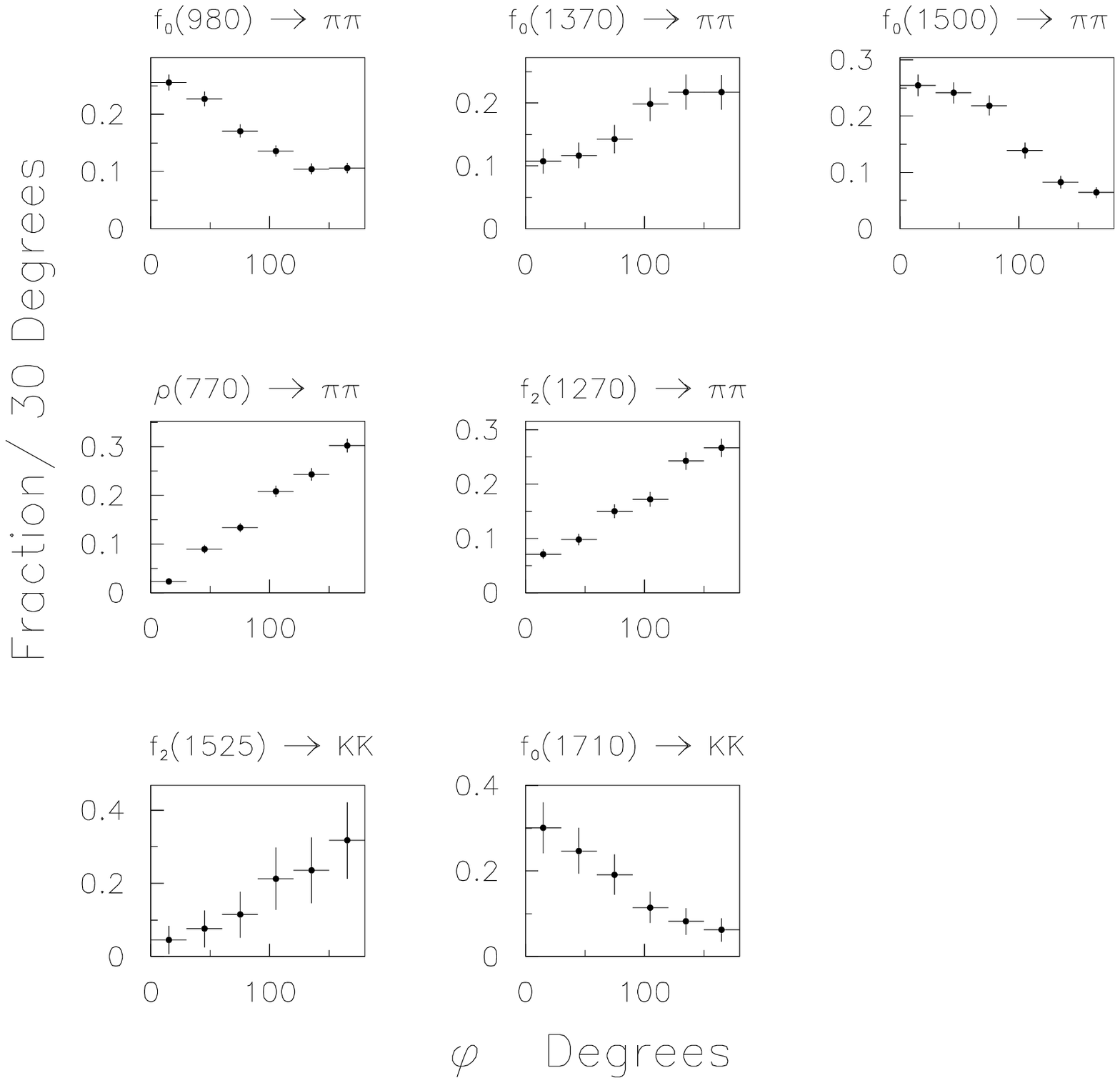,height=22cm,width=17cm,
bbllx=0pt,bblly=0pt,bburx=550pt,bbury=700pt}
\end{center}
\begin{center} {Figure 4} \end{center}
\newpage
\begin{center}
\epsfig{figure=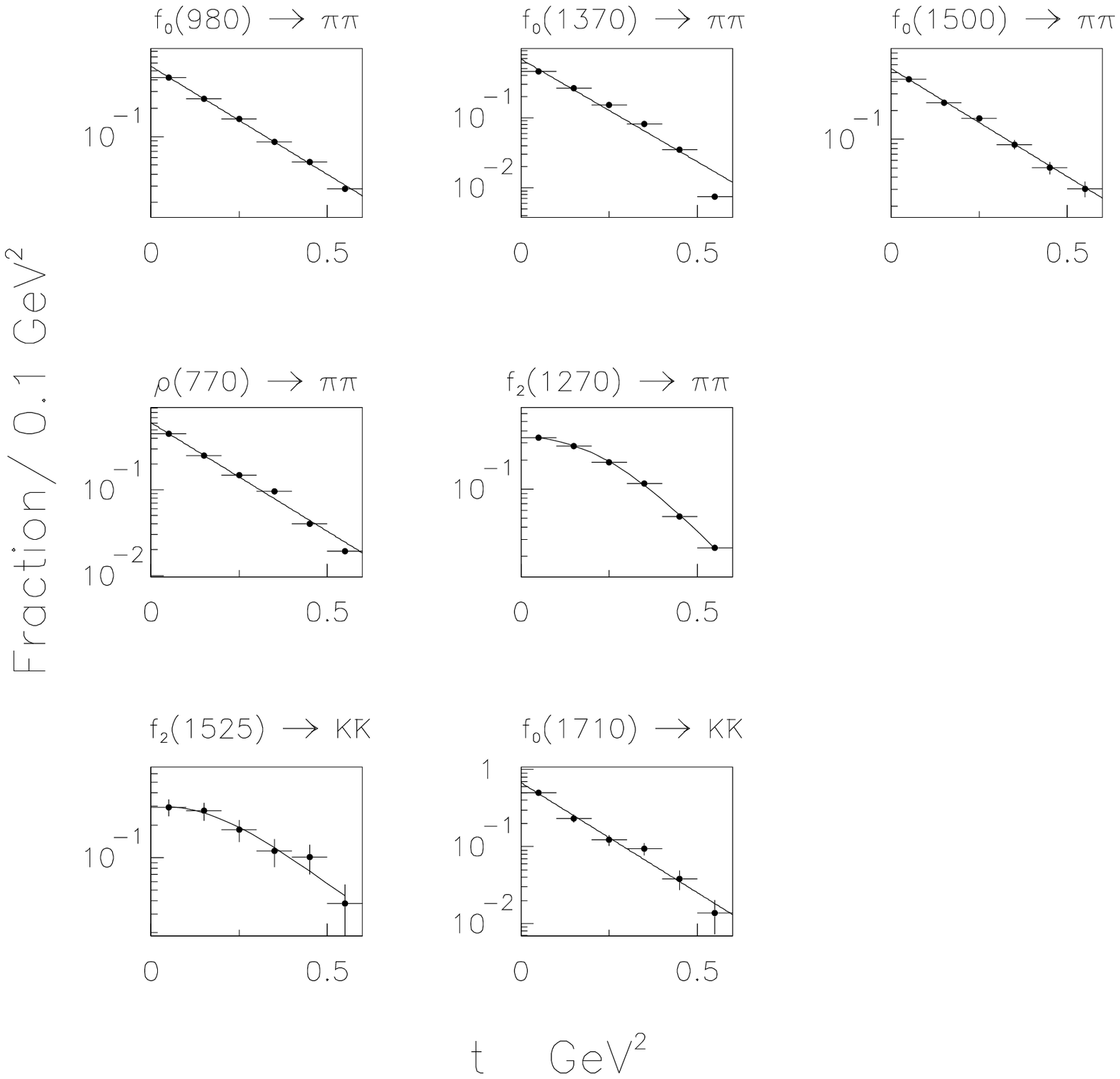,height=22cm,width=17cm,
bbllx=0pt,bblly=0pt,bburx=550pt,bbury=700pt}
\end{center}
\begin{center} {Figure 5} \end{center}
\end{document}